\documentclass[aip,jcp,reprint,floatfix,amsmath,amsfonts,10pt]{revtex4-1}
\usepackage{graphicx}
\usepackage{amsmath}
\usepackage[utf8]{inputenc}
\usepackage{color}
\usepackage{bbold}

\begin{document}

\title{{On the calculation of complete dissociation curves of closed-shell pseudo-onedimensional systems through the multireference method of increments}}
\author{E. Fertitta}
\author{B. Paulus}
\affiliation{Institut f\"ur Chemie und Biochemie - Takustr. 3, 14195 Berlin,\\ Freie Universit\"at Berlin, Germany}
\author{G. Barcza}
\author{\"O. Legeza}
\affiliation{Strongly correlated systems "Lend\"ulet" research group,\\ Wigner Research Centre for Physics, P.O.Box 49 Hungary}

\date\today

\begin{abstract}
The Method of Increments (MoI) has been employed using a multireference approach to calculate the dissociation curve of beryllium ring-shaped clusters Be$_n$ of different sizes. Benchmarks obtained through different single  and multireference methods including the ab initio Density Matrix Renormalization Group (DMRG) were used to verify the validity of the MoI truncation which showed a reliable behavior for the whole dissociation curve. Moreover we investigated the size dependence of the correlation energy at different distances in order to extrapolate the values for the periodic chain and to discuss the transition from a metal-like to a insulating-like behavior of the wave function through quantum chemical considerations.
\end{abstract}

\maketitle

\vspace{1cm}
\section{Introduction}

Metal-insulator transitions (MIT)\cite{MIT1,MIT2,MIT3,MIT4,MIT5} in realistic systems are often considered in a one-particle picture to explain the change of conductivity in a material. In situations such as band, Peierls and Anderson insulators, this yields a successful description, but in the case of Mott insulators the electron-electron correlation gives rise to a contribution which is more important than the electron-ion interaction in the localization of the electronic wave function (WF)\cite{Gebhard, Mott}. In this situations a proper description of the correlated WF becomes crucial, which cannot be in general achieved through the first principle methods usually employed for periodic systems such as the Density Functional Theory (DFT)\cite{DFT1,DFT2,DFT3}. Among the different wave function-based local approaches employed to overcome the dimensionality problem in correlated extended and periodic systems, the method of increments (MoI)\cite{paulus2003, paulus2006, paulus2007, stoll2009, Staemmler2009, mueller2011, voloshina2012} has gained a particular attention in the last decade. This approach can be applied to any wave function method and as shown in the recent work by Voloshina and Paulus\cite{voloshina2014} it successfully retrieves almost 100\% of the correlation energy of bulk materials with large static correlation if applied in a multireference (MR) fashion.\\
In our previous investigation\cite{dmrg_mit} we have exploited the quantum chemical version\cite{chan-rev,reiher-rev,yanai-rev,wooters-rev,szalay-rev} of the density matrix renormalization group (DMRG)\cite{white92,white93} approach to calculate the ground state energy of a model system, \emph{i.e.} beryllium ring-shaped cluster, and explored the use of quantum information theory (QIT)\cite{legeza2003, legeza2004, legeza2006-qpt, rissler06, barcza2010a, boguslawski2012b, boguslawski2013a} to characterize the wave function and thus determine the metal-like and insulating-like character of a system in different regions of the potential energy surface (PES). In the present paper we will focus again on the same system exploring the use of the method of increments (MoI) for closed-shell systems obtaining whole ground state dissociation curves through a multireference approach that allows to describe the crossing region where single reference approaches such as CCSD(T)\cite{CCSD1,CCSD2,CCSD3,CCSD4} fail. DMRG calculations as described in our previous work will be used as reference for testing the MoI approximation to the correlation energy.\\
Even though a model system is considered, we will underline how the use of standard canonical methods become prohibitive because of the high correlation effects involved especially if aiming to the thermodynamic limit. As we will show the MoI formalism used in this work involves only localized orbitals (LOs) which allows it to use limited active space despite large systems are involved. We will exploit this tool to calculate the correlation energy of  Be$_n$ rings as large as $n=90$ and extrapolate the behavior at the thermodynamic limit.\\
This paper is structured as follows: in section~\ref{sec:MoI} we describe the formalism of the method of increments for single- and multireference methods as applied in this work; in section~\ref{sec:comp_det} the problematics in describing the system with canonical wave function methods are underlined and we give the details of our calculations; in section~\ref{sec:results} we focus on th results obtained for the Be$_6$ ring in order to highlight the accuracy and the advantages of the multireference MoI in comparison different methods; we also report and compare the results obtained for larger rings and the behavior at the thermodynamic limit is extrapolated; our conclusions are finally drawn in section~\ref{sec:conclusion}.

\section{The Method of Increments}\label{sec:MoI}

\subsection{General Formalism}

The method of increments exploits the short range nature of the electronic correlation. Within this approach localized orbitals are used in order to describe the correlation energy as sum of individual contributions coming from the correlation of different parts of the systems, to which we will refer as bodies. In its general formalism which employs single reference methods such as coupled cluster (CC) or perturbation theory approaches, only the occupied orbitals are localized and some of them together with the virtual canonical orbitals are used to build the correlation space.\\
In the MoI, one starts from a first approximation of the correlation energy, $E^{\rm I}_{\rm corr}$, which is given by the sum of all independent contributions $\epsilon_{i}$ each arising from the $i^{th}$ body:
\begin{equation}
    E^{\rm I}_{\rm corr} = \sum_i \epsilon_{i}\label{eq_1body}
\end{equation}
We refer to the individual terms $\epsilon_{i}$ as 1-body increments. The contribution expressed in Eq. \ref{eq_1body} ranges typically between 60\% and 90\% of the correlation energy achievable through the chosen correlation method. The remaining part of $E_{\rm corr}$ is enclosed in the higher order increments which consider the correlation among many bodies. The second natural step is then to include one more body and define the 2-body increments $\Delta \epsilon_{ij}$ as:
\begin{equation}
\Delta \epsilon_{ij} = \epsilon_{ij} - \left(\epsilon_i + \epsilon_j\right)
\end{equation}
Going forward to higher order increments we get similar expressions. For instance the 3-body increments are calculated using the expression:
\begin{equation}
\Delta \epsilon_{ijk} = \epsilon_{ijk} - \left(\Delta \epsilon_{ij} + \Delta \epsilon_{jk} + \Delta \epsilon_{ik}\right) - \left(\epsilon_i + \epsilon_j + \epsilon_k\right)
\end{equation}
Combining these contributions, one can finally express the correlation energy $E_{\rm corr}$:
\begin{equation}
E_{\rm corr} = \sum_i \epsilon_i + \sum_{i<j} \Delta \epsilon_{ij} + \sum_{i<j<k} \Delta \epsilon_{ijk} + \cdots\label{ecorr_sum}
\end{equation}
As pointed out above, the electronic correlation is in general short ranged, and the $ee$-interaction is a two-particle interaction which implies that the increments decrease with the distance and the order of the bodies, \emph{i.e.} if the following convergence criteria are fulfilled:
\begin{equation}
|\Delta \epsilon_{ij}| > |\Delta \epsilon_{ik}| \mbox{ \hspace{0.1cm}  for \hspace{0.1cm} } r_{ij} > r_{ik}\label{conv1}
\end{equation}
\begin{equation}
|\Delta \epsilon_{ij}| > |\Delta \epsilon_{ijk}| > |\Delta \epsilon_{ijkl}| > \cdots\label{conv2}
\end{equation}
This allows to truncate the expansion of Eq.~\ref{ecorr_sum}, yielding a meaningful amount of the correlation energy. The local nature of the MoI and the possibility of truncating Eq.~\ref{ecorr_sum} are crucial points in making the method a candidate for the application to extended and periodic systems, for which canonical wave function methods are generally prohibitive. Nevertheless the MoI cannot be universally applied and fails when the required convergences explicated in Eq.~\ref{conv1}~and~\ref{conv2} do not occur. Moreover since the increments might have alternating sign, the MoI is not a variational method.

\subsection{Multireference Formalism}

\begin{figure}[htb]
\centerline{
\includegraphics[width=0.45\textwidth]{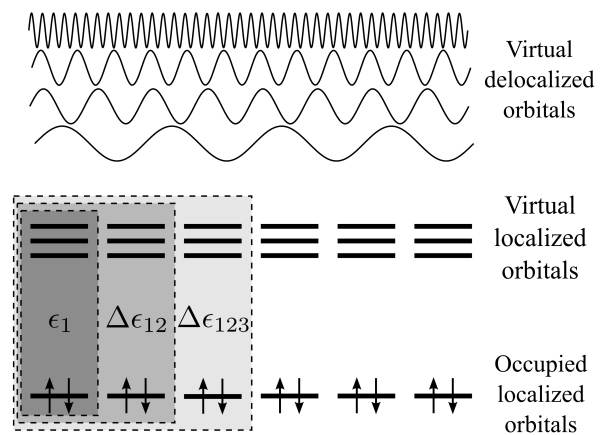}
}
\caption{(Color online) Sketch of the partition of the active orbitals for the method of increments. In its multireference formalism, CAS-SCF calculations are performed using localized orbitals only as shown in the highlighted boxes in order to obtain the non-dynamical incremental contributions to the correlation energy. On top of the obtained CAS wave functions, multireference calculations can be performed including excitation to the delocalized virtuals.} 
\label{fig:incre_scheme}
\end{figure}

As recently shown by Voloshina and Paulus \cite{voloshina2014}, a multireference MoI approach can successfully be used to calculate the cohesive energy of bulk metals for systems with a high static correlation contribution. In comparison with the single reference formalism, a different localization pattern is required in this approach. Indeed, besides the occupied orbitals, also the virtual orbitals important for the evaluation of the static correlation have to be localized. This way one can then calculate incremental static contributions to $E_{\rm corr}$ arising from Complete Active Space Self Consistent Field (CAS-SCF)\cite{ref_CASSCF} calculations performed within the bodies constituted by occupied and virtual LOs. Finally on top of the CAS-SCF wave function, a MR calculation is performed for each term of the incremental expansion, including the remaining delocalized virtual orbitals. The scheme is sketched in Fig.~\ref{fig:incre_scheme}. The equations described in the previous section remain still valid are used to expand and truncate the correlation energy. As described in the following section we will employ a minimal basis set and localize all virtual orbitals which will be use to construct the complete active space. We will not perform any multireference calculation on top of the CAS-SCF wave functions and we will refer to the method as CAS-MoI.

\section{Calculation details}\label{sec:comp_det}
\subsection{The System}

\begin{figure}[htb]
\centerline{
\includegraphics[width=0.45\textwidth]{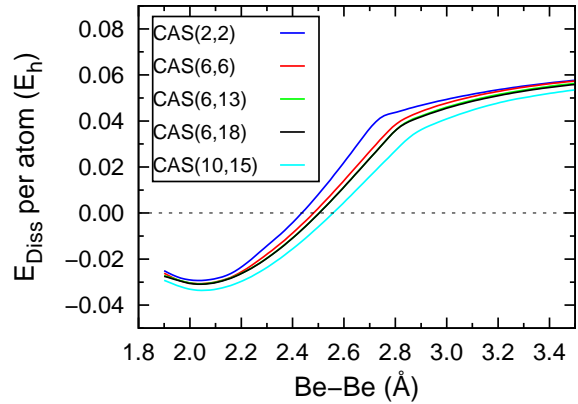}
}
\caption{(Color online) Dissociation curves calculated for the ground state of Be$_{10}$ at the Complete Active Space Self Consistent Field (CAS-SCF) level of theory, using different active spaces. The dashed line corresponds to the dissociation limit calculated using a CAS(2,4) for the Be atom. A minimal basis set has been used in all cases.}
\label{fig:be10_cas}
\end{figure}

\begin{figure}[htb]
\centerline{
\includegraphics[width=0.45\textwidth]{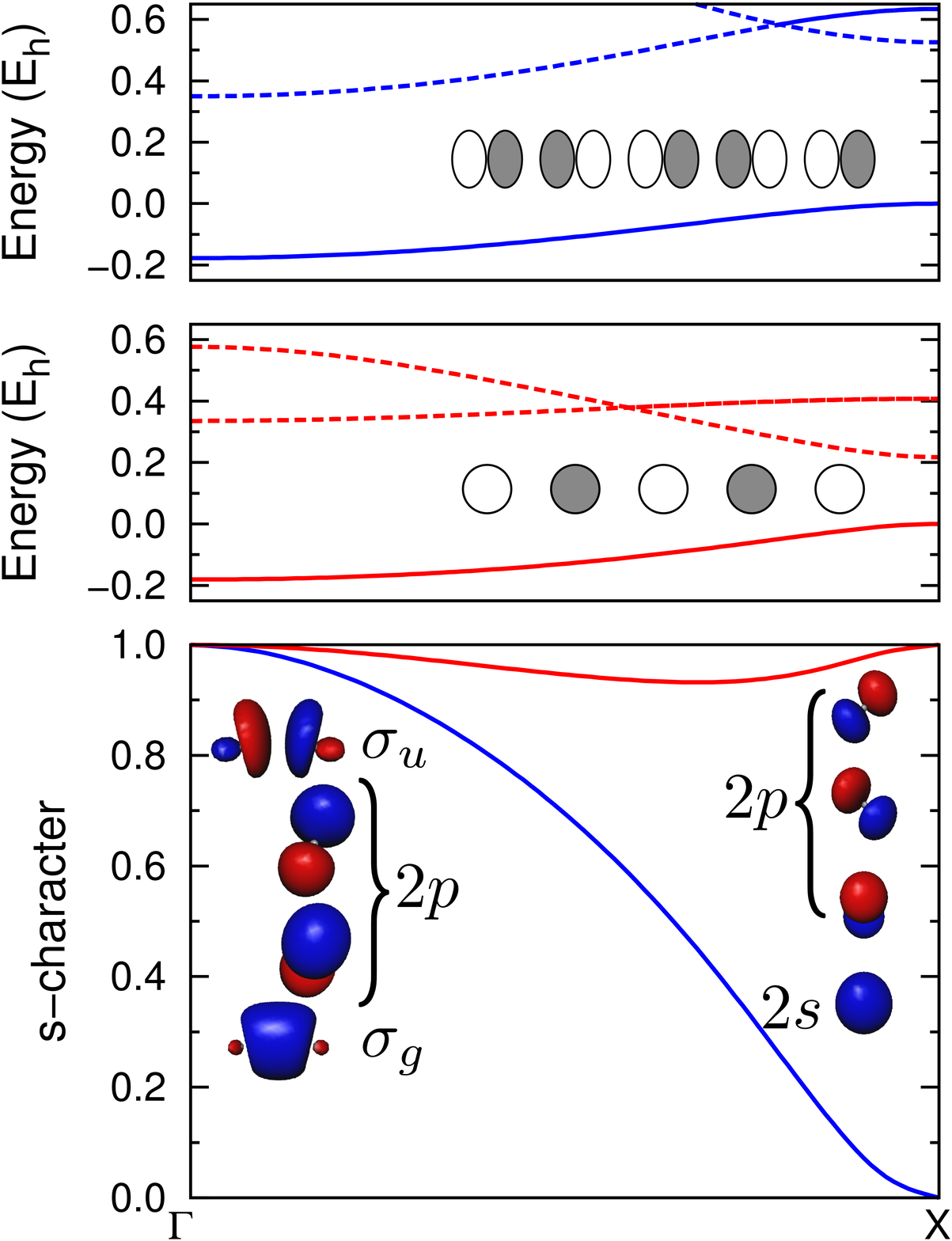}
}
\caption{(Color online) Hartree-Fock band structure and $s$-character of the valence crystal orbitals of an infinite beryllium chain for the two leading configurations Conf~1 (blue line) and Conf~2 (red line). The $s$-character, defined as the normalized sum of the squared coefficients of the $s$-bases, is reported in reciprocal space as well as the valence band for the two regimes. Notice that the character of the valence band (full lines) at the X-point of the Brilluoin zone switches from pure $p$ to pure $s$ as sketched in the insets. The localized orbitals obtained by unitary transformation of the valence and virtual orbitals are also shown. As one can see, the difference in $s$-character is reflected by the different symmetry of the localized orbitals.
}
\label{fig:configurations}
\end{figure}

As subject of our investigation we have chosen a model system characterized by a high static correlation contribution, \emph{i.e.}~beryllium ring-shaped clusters, whose pseudo-onedimensional structure was chosen to resemble the periodic Born-von-Karman boundary conditions and to obtain a WF that will inevitably converge with size towards the thermodynamic limit of an infinite linear chain. Because of the quasi degeneracy of valence 2$s$ and virtual 2$p$ orbitals (more than 93\% of the correlation energy of Be atom in its $^1S$ ground state is static), 4$n$ active orbitals would be required to obtain a size-consistent CAS-SCF reference for MR calculations for Be$_n$. This leads to CAS(2$n$, 4$n$) calculations which become of course prohibitive as the number of Be atoms increases. We report in Fig.~\ref{fig:be10_cas} the dissociation curves of Be$_{10}$ calculated using a minimal atomic basis set and different active spaces within the CAS-SCF method. As one can see, despite the use of larger and larger active spaces gives a finer and finer description of the potential energy curves, the calculated dissociation plateaus lie at much higher energies than the dissociation limit calculated within the same method using an active space consisting of the 2$s$ and 2$p$ functions of the free atom. Similar results are reported in more details in Table~\ref{tab_ci} as described later on.\\
This model system is complicated further by the fact that the $p$ functions, besides enlarging the active space, play a big role at the Hartree-Fock level too. Indeed, while at dissociation the HF orbitals will be linear combination of pure 2$s$ atomic orbitals, as the interatomic distance shortens their $p$-character increases and, at a certain geometry, the HOMO switches from a pure $s$ to a pure $p$ molecular orbital. In other words, we encounter a crossing between two HF configurations which dominate the ground state at different interactomic distances. This holds for the finite pseudo-onedimensional clusters indepentely from their size as well for the periodic chain given that both the energy and the character of the discrete Hartree-Fock molecular orbitals of Be$_n$ rings converge towards the crystal orbitals. In order to supplye a graphical representation of these HF wave functions, we report in Fig.~\ref{fig:configurations} the band structure calculated in the two regimes and the $s$-character of the Hartree-Fock valence orbitals of a periodic beryllium chain close to the minimum of the dissociation curve (blue line) and towards dissociation (red line). Respectively, we will refer to these as configuration (Conf) 1 and 2 and we will indicate the two regimes where each of them dominates as metal-like and insulator-like regime. Clearly an accurate single reference method such as CCSD(T) will work perfectly in the regimes where one of the configuration is dominant, but not around the crossing region. We will show how the size problem and the necessity for a MR approach will be overcome through the use of the multireference method of increments allowing us to describe the PES of Be$_n$ till dissociation.\\
By unitary transformation of the canonical orbitals of the two main HF configurations different sets of LOs are obtained (see Fig.~\ref{fig:configurations}). Indeed, in the insulator-like regime localized orbitals resemble atomic 2$s$ and 2$p$ orbitals, while in the metallic regime $\sigma_g$- and $\sigma_u$-like orbitals appear. As we will show, the use of the different starting Hartree-Fock configuration has a huge impact on the effectiveness of the method, despite in both cases the result should converge toward the Full-CI limit.

\subsection{Computation and basis set}

The MOLPRO quantum chemistry package\cite{MOLPRO} was used to perform the different steps of MoI calculations {\emph i.e.} Hartree-Fock calculations, Foster-Boys localization\cite{Boys} and CAS-SCF calculations. We employed a minimal basis set consisting of 1$s$, 2$s$ and 2$p$ atomic functions derived by the Dunning's $cc$-pVDZ\cite{cc_pVDZ}. We kept the core 1$s$ orbitals frozen during the incremental calculations, focusing on the correlation of valence orbitals. The Crystal09\cite{crystal} code was employed to calculate the Hartree-Fock wave function of the periodic chain using the same basis set.

\section{Results}\label{sec:results}

\subsection{Be$_6$ ring}

\begin{figure}[htb]
\centerline{
\includegraphics[width=0.45\textwidth]{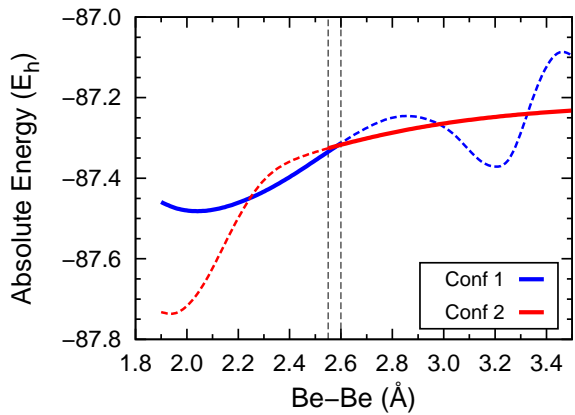}
}
\caption{(Color online) Potential energy curve for Be$_6$ calculated using the MoI at the 4-body level within the CAS-SCF approach starting from the two HF main configurations.}
\label{fig:PES_MR-MoI}
\end{figure}

\begin{figure}[htb]
\centerline{
\includegraphics[width=0.45\textwidth]{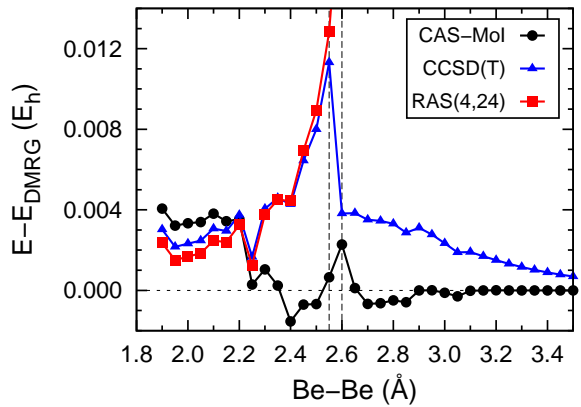}
}
\caption{(Color online) Comparison between the differences of the PES for Be$_6$ obtained through different methods and DMRG($M=1024$).}
\label{fig:diff_DMRG}
\end{figure}

\begin{figure*}[htb]
\centerline{
\includegraphics[width=0.9\textwidth]{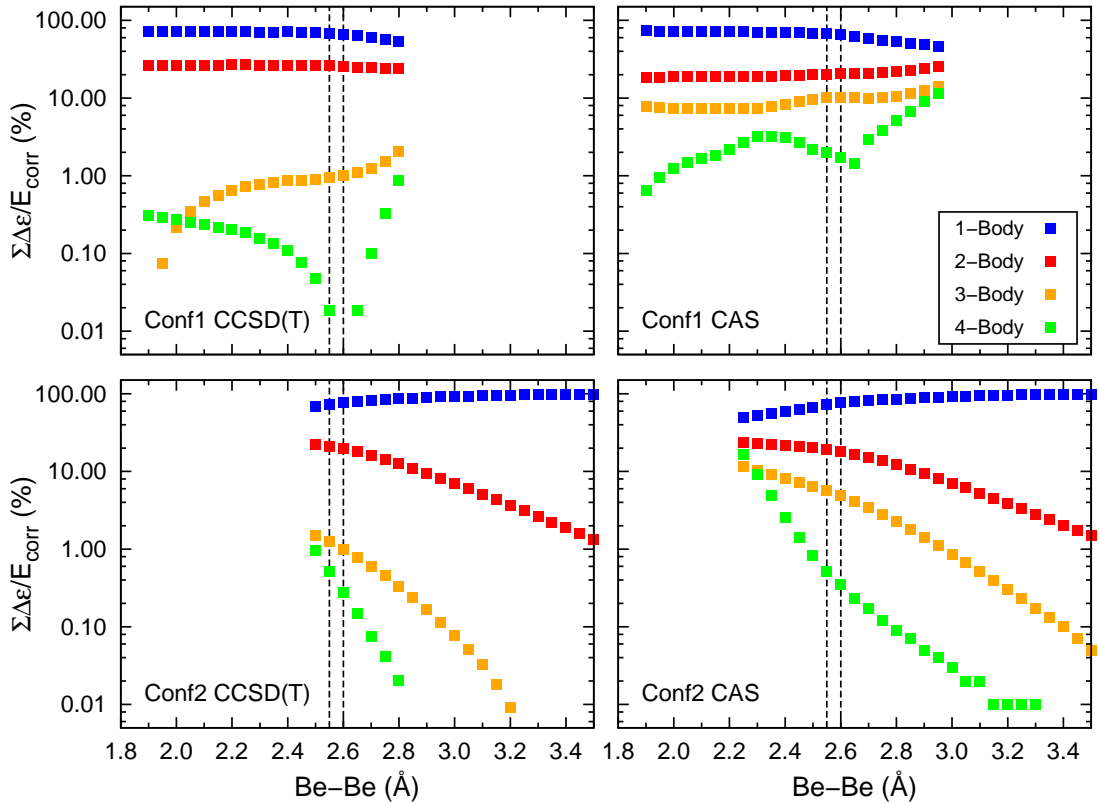}
}
\caption{(Color online) Percent of the correlation energy contribution of every incremental order as a function of distance for a Be$_6$ ring. The data are reported for configuration 1 (doubly occupied $sp$ hybrids) and configuration 2 (2$s^2$). The convergence criteria fulfill only in particular distance regimes where the incremental scheme can successfully be used and more than 99\% of the correlation energy can be retrieved at the 4-body level. At the crossing point 2.60\AA, both configurations can be used.}\label{fig:incre_percent}
\end{figure*}

\begin{figure}[htb]
\centerline{
\includegraphics[width=0.45\textwidth]{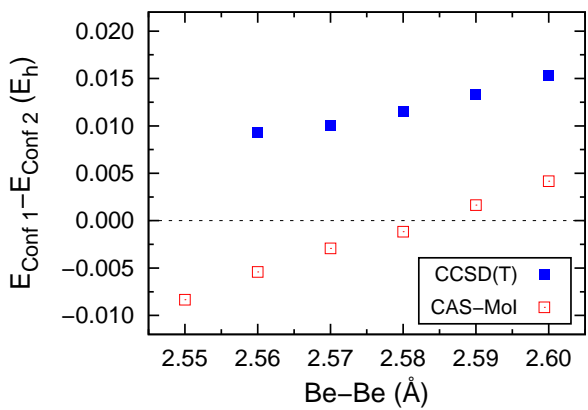}
}
\caption{(Color online) Comparison between the energy difference $E_{\rm Conf 1}-E_{\rm Conf 2}$ for the canonical CCSD(T) and the CAS-MoI around the crossing for Be$_6$. At the Full-CI level this difference is expected to be zero. At the 4-body level the method of increments gives an acceptable result.}\label{fig:diff_conf1_conf2}
\end{figure}

We start our analysis reporting the single reference CCSD(T) results obtained for Be$_6$ using both the canonical formalism and the MoI expansion, to illustrate how within the same quantum chemical method, this truncation works in retrieving the correlation energy. The method of increments at the 4-body level retrieves more than the 99.99\% of the correlation energy achievable with the canonical CCSD(T) if the proper HF starting configurations (Conf~1~and~2) are used in the two limiting regimes, while around the crossing this percentage drops to around 95\%. Also in this regime the $\mathcal T_1$ diagnostic calculated within the canonical CCSD(T) presents values larger than $0.03$ highlighting the necessity of using a multireference approach for describing the static correlation. However as already stated and shown in Fig.~\ref{fig:be10_cas}, canonical CAS-SCF approaches cannot be used to calculate a size consistent wave function unless a CAS(2$n$,4$n$) is performed, \emph{i.e.} a Full-CI in our minimal basis set, and this is of course not applicable as $n$ increases. Of course, a multireference calculation performed on top of an accurate CAS or RAS calculation would reach the required accuracy, but since it is our interest to explore the convergence toward the thermodynamic limit, it is clear that a local method is preferable. Moreover, this choice in general reduce the problem of choosing a proper and consistent active space.\\
In Fig.~\ref{fig:PES_MR-MoI} we report the dissociation curve of Be$_6$ as obtained using the CAS-MoI at the 4-body level using the Conf~1~and~2. Because in our case, the incremental scheme is converging towards the Full-CI solution, it would be in principle equivalent to start from one or the other configuration. However as it can be seen, truncating the correlation energy at the 4-body level the two results match in a narrow regime only, \emph{i.e.} around the crossing, while for other internuclear distances a non-monotonic behavior is obtained if not the proper HF configuration is used. We want to focus now on the comparison with the DMRG data which were obtained in our previous investigation\cite{dmrg_mit} and we use fixed $M=1024$ block states results as a reference to evaluate the behavior of the CAS-MoI and the other methods discussed. In Fig.~\ref{fig:diff_DMRG} we report the absolute error with respect to the DMRG data of the above described CAS-MoI, canonical CCSD(T) and the very accurate multiconfigurational calculation, RAS(4,24). As it can be seen the CAS-MoI gives the best results for almost any interatomic distance, and while the other approaches depart too much from the DMRG reference around the crossing, the CAS-MoI results stay within an error of $\pm~4~{\rm m}E_h$ on the total energy. Moreover since the increments are all negative and converging  (see Fig.~\ref{fig:incre_percent}) we might argue that the MoI is behaving variationally in our calculations. As a consequence, if this is the case, the method of increments is actually retrieving more correlation energy than DMRG (as one can see the error is negative in a quite wide regime). Only in the metal-like regime this approach gives worse results with respect to the other reported even if it underestimate the module of $E_{\rm corr}$ only by $2\times10^{-3}~E_h$ with respect to RAS(4,24)). This can be explained considering the high delocalization of the wave function in this regime which is hard to describe with a local approach. A quantitative comparison between the results with CAS-MoI, DMRG, CCSD(T) and different CAS and RAS methods is reported in Table~\ref{tab_ci}. Again one can realize the difficulty in retrieving a meaningful amount of the correlation energy with canonical methods and how in general the local approach gives reasonable results with relatively less effort.\\
Let us now consider the strong deviations shown in Fig.~\ref{fig:PES_MR-MoI} (dashed lines) that occur when the CAS-MoI is employed starting from a HF configuration which is not the dominant one. In order to explain why in this situation the method fails in describing the electronic structure of the system, let us analyze the behavior of the individual increments as shown in Fig.~\ref{fig:incre_percent}. What can be observed is that, as we overcome the crossing, the required convergence stated in Eq.~\ref{conv1}~and~\ref{conv2} break down and higher order increments give comparable contributions to $E_{\rm corr}$ as lower order ones. This is equivalent to say that in the employed orbital basis (\emph{i.e} HF configuration), higher order contributions become more and more important which is analogous to what was concluded from the analysis of the mutual information as described in our DMRG analysis\cite{dmrg_mit} that showed the increase of long range entanglement. This strong dependence on the starting configuration is of course a great disadvantage of the MoI which can otherwise yield  accurate results with relatively cheap calculations.\\
Since we are forced to calculate both configurations, it might appear that no particular advantage arises from the use the CAS-MoI with respect to the single reference approach. On account of this, in Fig.~\ref{fig:diff_conf1_conf2} we report the differences between the energies obtained from Conf~1 and Conf~2 in the crossing regime for both methods. As it can be seen, in the case of CAS-MoI this difference is around $10~{\rm m}E_h$ smaller than CCSD(T) and ranges between $\pm5~{\rm m}E_h$. This means that in this narrow regime, the two bases give us comparable results leading to a smooth curve, which is not the case for CCSD(T). In account of this and previous statements, we can conclude that the CAS-MoI allows to describe the behavior of the PES also in the crossing regime allowing us to describe the whole dissociation curve.\\
We conclude this section comparing the individual increments as obtained from CCSD(T) and CAS-SCF (see Fig.~\ref{fig:incre_percent} and Table~\ref{tab_incre}). For the former method, the lower order increments are larger than for CAS-MoI because a larger active space is involved, but at the 3- and 4-body level the orbital optimization which retrieves the static correlation starts playing a bigger role than the size of active space. In general the single reference method converges faster, but the total correlation energy is larger for CAS-SCF than for the single reference approach.
\begin{table*}[hb]\fontsize{9}{9}\selectfont
\centering
\caption{(Color online) Comparison between different canonical CAS and RAS approaches, CCSD(T), DMRG calculations reported in Ref.[\onlinecite{dmrg_mit}] and MoI calculation at the 4-body level as described in the present paper for Be$_6$.}\label{tab_ci}
\begin{tabular}{lrccp{1.5cm}rcc}
                     & \multicolumn{3}{c}{Metal-like -- 2.10\AA -- Conf 1}          && \multicolumn{3}{c}{Insulator-like -- 3.00\AA -- Conf 2}  \\
\cline{2-4}\cline{6-8}
Calculation          &     CSFs &   Energy ($E_h$) & $E_{\rm corr} (\%)$ &&     CSFs &   Energy ($E_h$) & $E_{\rm corr} (\%)$ \\
\hline                                                                                                  
CAS(6,6)         &         34&	-87.329677	&        8.09\%	&&         34&	-86.947385	&        8.54\%	\\
CAS(6,12)        &	2,086&	-87.364902	&       29.31\%	&&	2,086&	-87.034032	&	33.57\%	\\
CAS(6,15)        &	8,155&	-87.367285	&       30.74\%	&&	8,155&	-87.035797	&	34.08\%	\\
RAS(4,21)        &     13,523&	-87.373196	&       34.30\%	&&     13,521&	-87.002608	&	24.49\%	\\
CAS(6,21)        &     64,835&	-87.373263	&       34.34\%	&&     64,835&	-87.002788	&	24.54\%	\\
RAS(6,18)        &  1,268,308&	-87.425535	&       65.82\%	&&  1,268,308&	-87.146271	&	65.98\%	\\
RAS(4,23)        &    134,843&	-87.428804	&       67.79\%	&&    134,837&	-87.141082	&	64.48\%	\\
RAS(2,24)        &        884&	-87.455551	&       83.90\%	&&        885&	-87.160246	&	70.02\%	\\
CCSD(T)	         &           & 	-87.479226	&       98.15\%	&&           &	-87.261723	&	99.32\%	\\
RAS(4,24)        &    295,746&	-87.479826	&       98.51\%	&&    295,752&	-87.240226	&	93.12\%	\\
RAS(6,24)        & 14,972,954&	-87.482727	&      100.26\%	&& 14,972,954&	-87.260875	&	99.08\%	\\
\hline											
CCSD(T)-MoI(4-body)&         &	-87.479272	&	98.18\%	&&           &	-87.261705	&	99.32\%	\\
MR-MoI(4-body)     &  866,320&	-87.478599	&	97.77\%	&&    866,320&	-87.264183	&      100.03\%	\\
\hline											
DMRG($M=1024$)	&	&	-87.482294	&	100.00\% &&	&	-87.264065	&	100.00\%	
\end{tabular}
\end{table*}
\begin{table*}[h]\fontsize{9}{9}\selectfont
\centering
\caption{Individual incremental orders and total correlation energy of Be$_6$ calculated with the MoI at the 4-body level using a single reference method, CCSD(T) and the multireference method as described in this paper.}\label{tab_incre}
\begin{tabular}{lp{0.2cm}ccp{0.4cm}ccp{0.4cm}ccp{0.4cm}cc}
         & & \multicolumn{5}{c}{Configuration 1} && \multicolumn{5}{c}{Configuration 2}\\
\cline{3-7}\cline{9-13}
   & & \multicolumn{2}{c}{Metal-like -- 2.10\AA} && \multicolumn{2}{c}{Crossing regime -- 2.60\AA} && \multicolumn{2}{c}{Crossing regime -- 2.60\AA} && \multicolumn{2}{c}{Insulator-like -- 3.00\AA}  \\
\cline{3-4}\cline{6-7}\cline{9-10}\cline{12-13}
Increments                             &  &   CCSD(T)   &  CAS-SCF  &&  CCSD(T)     &  CAS-SCF    &&  CCSD(T)    &  CAS-SCF      &&    CCSD(T)   &  CAS-SCF \\
\hline                                                                                                 
$\sum_i \epsilon_i$                    &  &-0.0197077  &-0.0195083&&-0.0282556   &  -0.0278839&&-0.0443223  & -0.0438349   &&  -0.0532219  &-0.0531172\\
$\sum_{i,j} \Delta\epsilon_{ij}$       &  &-0.0072719  &-0.0051364&&-0.0109604   &  -0.0088279&&-0.0113016  & -0.0105205   &&  -0.0040539  &-0.0041013\\
$\sum_{i,j,k} \Delta\epsilon_{ijk}$    &  &-0.0001276  &-0.0019666&&-0.0004272   &  -0.0043662&&-0.0005745  & -0.0028508   &&  -0.0000447  &-0.0004985\\
$\sum_{i,j,k,l} \Delta\epsilon_{ijkl}$ &  &-0.0000652  &-0.0004490&&-0.0000009   &  -0.0007402&&-0.0001604  & -0.0001981   &&   0.0000035  &-0.0000157\\
\hline                                                                                                   
$E_{\rm corr}$                         &  &-0.0271724  &-0.0270603&&-0.0396440   &  -0.0418181&&-0.0563588  & -0.0574042   &&  -0.0573169  &-0.0577327\\
\vspace{-0.2cm}\\
$E_{\rm tot}$                          &  &-87.479272  &-87.478599&&-87.300283   & -87.3133272&&-87.310961  & -87.317234   &&  -87.261705  &-87.264200
\end{tabular}
\end{table*}

\subsection{Extension to larger rings}

\begin{figure}[htb]
\centerline{
\includegraphics[width=0.45\textwidth]{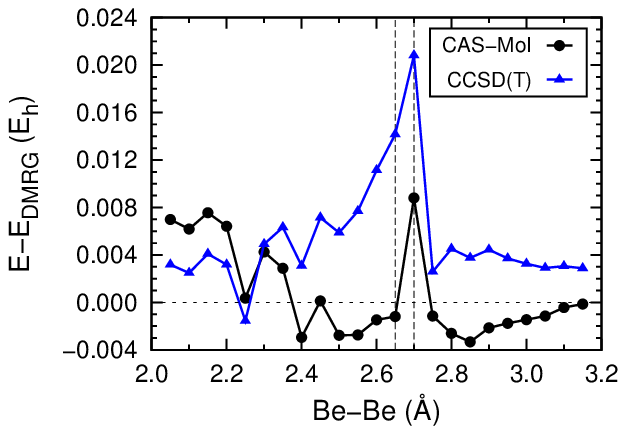}
}
\caption{(Color online) Energy difference between CAS-MoI at the 4-body level and DMRG($M=1024$) for the Be$_{10}$.}
\label{fig:diff_dmrg_10}
\end{figure}

\begin{figure}[htb]
\centerline{
\includegraphics[width=0.45\textwidth]{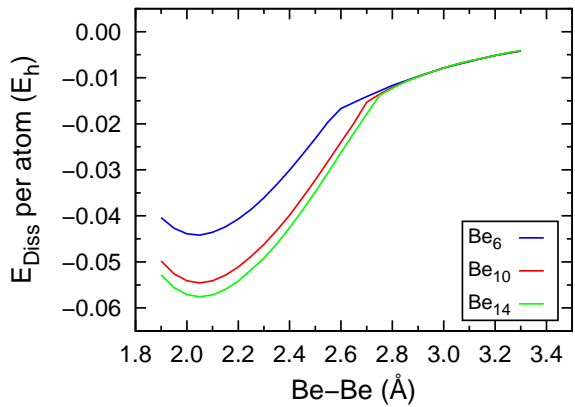}
}
\caption{(Color online) Potential energy surfaces for Be$_6$, Be$_{10}$ and Be$_{14}$ rings obtained through the CAS-MoI at the 4-body level.}
\label{fig:6_10_14}
\end{figure}

\begin{figure*}[htb]
Metal-like regime -- 2.10{\AA} Conf 1\\
\centerline{
\includegraphics[width=0.9\textwidth]{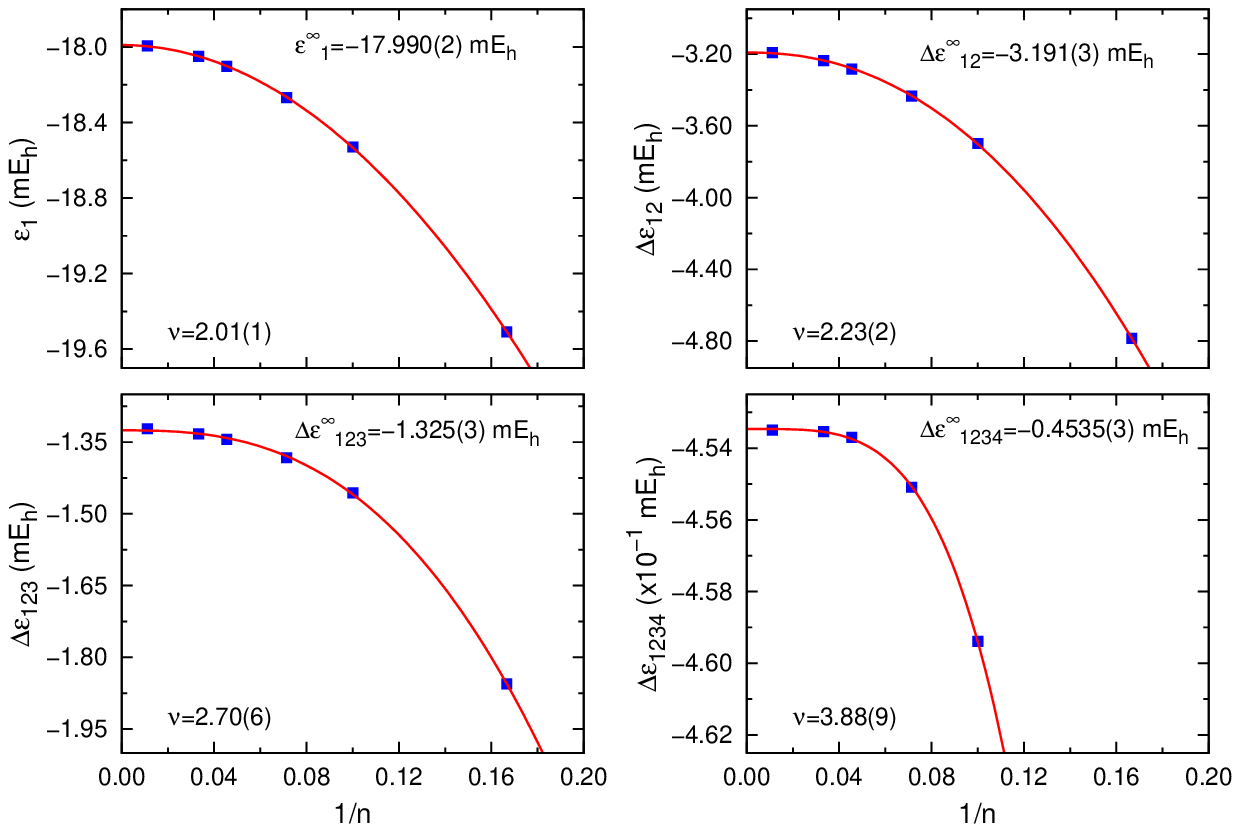}
}
Insulator-like regime -- 3.00{\AA} Conf 2\\
\centerline{
\includegraphics[width=0.9\textwidth]{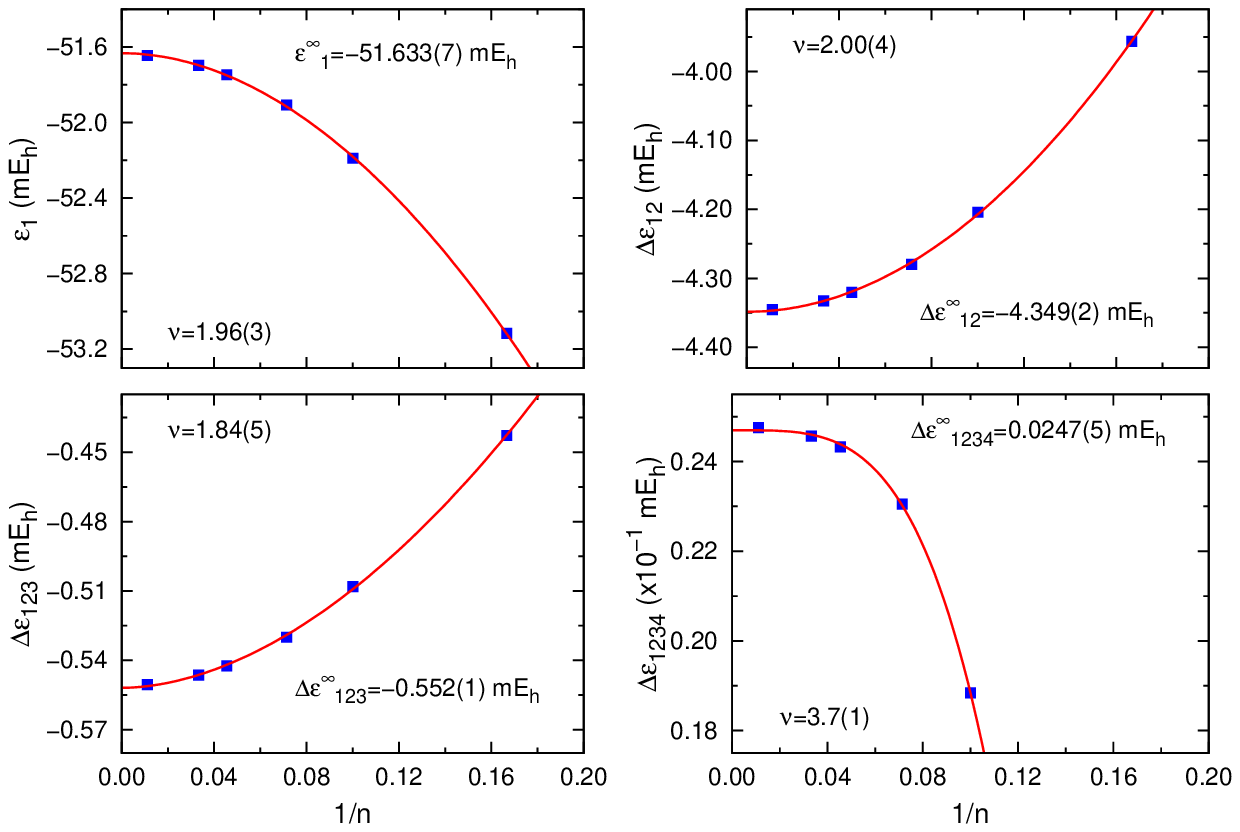}
}
\caption{(Color online) Convergence with the ring size of the largest individual increments of each order. The red lines were obtained by fitting the calculated data with an equation of the form $a/n^\nu+E^{\infty}$ where $E^{\infty}$ is the value extrapolated for the periodic chain.}\label{fig_fits}
\end{figure*}

\begin{figure*}[htb]
\centerline{
\includegraphics[width=0.9\textwidth]{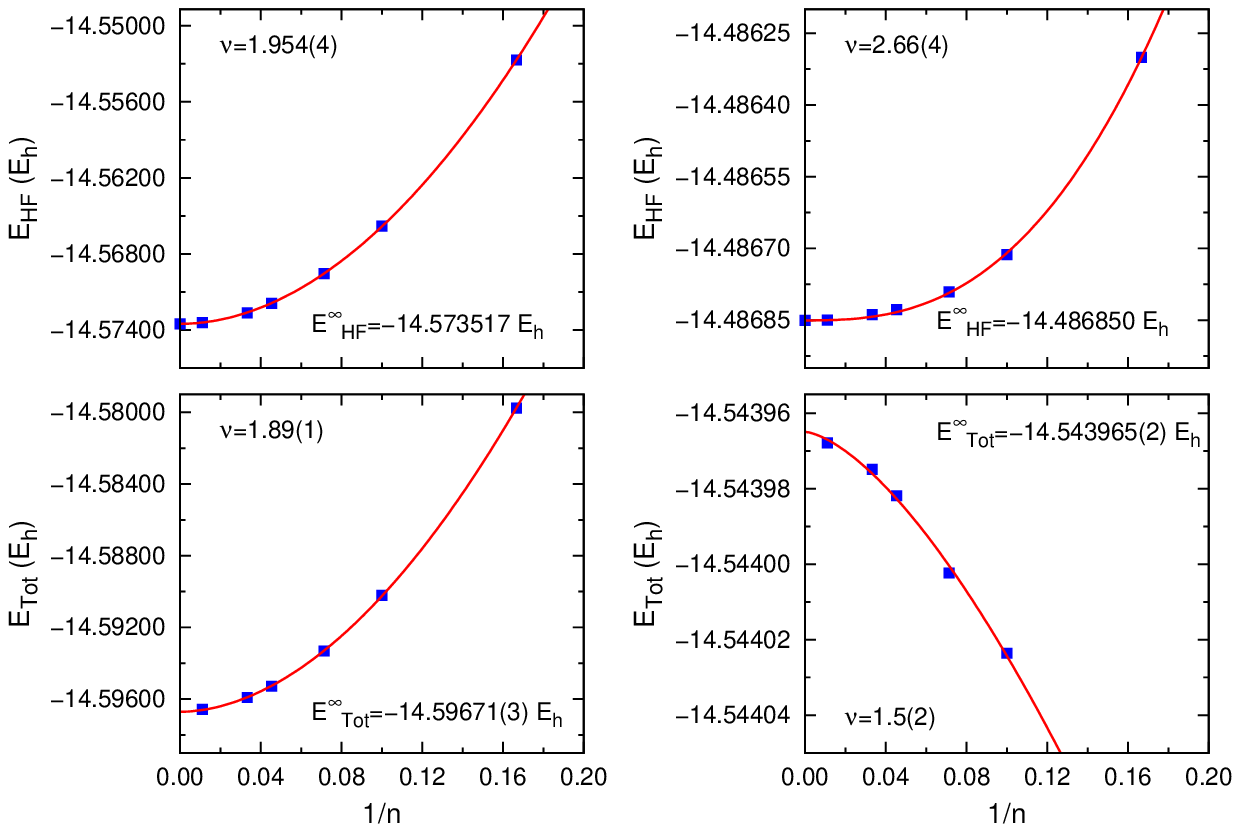}
}
\caption{(Color online) Convergence with the ring size of the Hartree-Fock and total energy per atom for 2.10\AA (left) and 3.00\AA (right).}\label{HF_trends}
\end{figure*}

After analyzing the use of the method for Be$_6$ we report the results for larger rings in order to evaluate the convergence towards the thermodynamic limit. DMRG allowed us to describe the dissociation of Be$_{10}$ (20 electrons in 40 active orbitals) and as for the smaller cluster we report (Fig.~\ref{fig:diff_dmrg_10}) the energy differences of CAS-MoI and CCSD(T) with respect to DMRG. Also in this case it can be seen that the former method behaves much better than the single reference one and that the obtained values differ from DMRG ones by around $5~{\rm m}E_h$.\\
As already stated, moving forward the aim of describing periodic systems, local methods become better candidate than canonical ones for obtaining accurate correlation energies. Within the CAS-MoI, the amount of necessary CSFs depends on the body order only and not on the size of the system, while in a canonical approach prohibitive active space sizes become necessary. The PES calculated for Be$_{10}$ and Be$_{14}$ rings at the 4-body level are shown in Fig.~\ref{fig:6_10_14} in comparison with Be$_6$. Even from these small cluster we can deduce the trend for increasing size, that is a faster convergence in the insulator-like regime than in the metal-like regime and a shift of the crossing towards larger internuclear distances.\\
In order to highlight these convergences we report in Table~\ref{tab_incre_size} the individual increments and the calculated correlation energies for Be$_n$ rings with $n=$ 6, 10, 14, 22, 30 and 90 in the two regimes. As one can see in Fig.~\ref{fig_fits} the individual increments can be fitted with a $n^{-\nu}$ function. This way through extrapolation it was possible to evaluate the values for $n\rightarrow\infty$, \emph{i.e.} for the periodic chain, which are also reported in Table~\ref{tab_incre_size}.\\
It can be observed that both in the metal-like and insulator-like regime the trend is basically quadratic for the 1-body while it rapidally deviates for higher order increments in the metal-like regime where the depencende on the angle is stronger. It has to be underlined that for the 4-body case, the value for $n=6$ was not included in the fit because for obvious geometrical reason was to much out of trend. Clearly the 1-body dominates the total correlation energy and therefore $E_{\rm corr}$ depends also quadratically on $1/n$. We conclude showing in Fig.~\ref{HF_trends} that a similar behavior was obtained for the Hartree-Fock energy too.\\
Of course a faster convergence would be achieved if linear chains had been considered since no angle dependence would be present as in the ring shaped cluster, but these were chosen to impose a higher symmetry to the system, deleting any border effect and allowing the different bodies to be equivalent to each other. Moreover the cyclic structure recalls more the periodic Born-von-Karman boundary conditions imposed in the periodic case.
\begin{table*}\fontsize{9}{9}\selectfont
\centering
\caption{Individual incremental orders calculated with the CAS-MoI for Be$_n$ ring-shaped cluster of different sizes form $n=$ 6 to 90 for two internuclear distances, 2.10{\AA} and 3.00{\AA}. The extrapolated values for the infinite chains are also reported.}\label{tab_incre_size}
Metal-like regime -- 2.10{\AA} Conf 1\\
\begin{tabular}{lccccccc}
    Increments                       &      6       &      10       &       14      &     22       &    30        &     90        & $n\rightarrow\infty$ \\
\hline
$\sum_i \epsilon_i$                    & -0.01950832  &  -0.01852919  &  -0.01826832  & -0.01810244  & -0.01804935  &   -0.01799396 &  -0.017990(2) \\
$\sum_{i,j} \Delta\epsilon_{ij}$       & -0.00513636  &  -0.00388669  &  -0.00358258  & -0.00340908  & -0.00335721  &   -0.00330521 &  -0.003305(3) \\
$\sum_{i,j,k} \Delta\epsilon_{ijk}$    & -0.00196657  &  -0.00152785  &  -0.00145735  & -0.00141879  & -0.00140766  &   -0.00139659 &  -0.001402(4) \\
$\sum_{i,j,k,l} \Delta\epsilon_{ijkl}$ & -0.00044904  &  -0.00048085  &  -0.00047395  & -0.00047172  & -0.00047140  &   -0.00047112 &  -0.00047118(4) \\
\hline
$E_{\rm corr}$                         & -0.02706030  &  -0.02442458  &  -0.02378220  & -0.02340203  & -0.02328562  &  	-0.02316688 &  -0.023155(2)
\end{tabular}\\
\vspace{0.6cm}
Insulator-like regime -- 3.00{\AA} Conf 2\\
\begin{tabular}{lccccccc}
      Increments                       &      6       &      10       &       14      &     22       &    30        &     90        & $n\rightarrow\infty$ \\
\hline
$\sum_i \epsilon_i$                    & -0.05311719  & -0.05218930   & -0.05190679   & -0.05174649  & -0.05169682  &	-0.05164440 &   -0.051633(7) \\
$\sum_{i,j} \Delta\epsilon_{ij}$       & -0.00410130  &	-0.00450650   & -0.00464482   & -0.00472404  & -0.00474799  &	-0.00477369 &   -0.004783(6) \\
$\sum_{i,j,k} \Delta\epsilon_{ijk}$    & -0.00049846  &	-0.00063636   & -0.00068277   & -0.00070780  & -0.00071598  &   -0.00072474 &   -0.000727(2) \\
$\sum_{i,j,k,l} \Delta\epsilon_{ijkl}$ & -0.00001574  &  0.00002165   &  0.00002305   &  0.00002433  &  0.00002457  &    0.00002476 &    0.0000247(5) \\
\hline
$E_{\rm corr}$                         & -0.05773268  & -0.05731051   & -0.05721133   & -0.05715400  & -0.05713622  &   -0.05711807 &   -0.057119(2)
\end{tabular}
\end{table*}

\section{Conclusion}\label{sec:conclusion}

The method of increments has been applied using a multireference formalism in order to describe the dissociation of ring shaped beryllium cluster of different sizes. The large static correlation involved in these systems made them good candidates for our analysis. Being size consistent and size extensive the MoI allowed us to obtain CAS-SCF wave function that can be further used as basis for multireference calculations. Key point of this investigation was that we were able to describe the whole PES also where single reference methods fail, \emph{i.e.} close to the crossing. DMRG calculations previously performed were crucial to evaluate the reliability of our results. The correlation energy of a system as large as Be$_{90}$ was obtained using this approach. We want to underline that within a canonical method this calculation would involve 180 active electrons in 360 active orbitals which without a local approach this would be impossible to treat. Finally, investigating different sizes of Be$_n$ rings we could evaluate the correlation energy for the periodic system.

\acknowledgments{This research was supported in part by the Hungarian Research Fund (OTKA) under Grants No. NN110360 and No. K100908, the Agence Nationale de la Recherche (ANR), and the German Research Foundation (DFG) via the project "Quantum-chemical investigation of the metal-insulator transition in realistic low-dimensional" (action ANR-11-INTB- 1009 MITLOW PA1360/6-1). The support of the Zentraleinrichtung f\"ur Datenverarbeitung (ZEDAT) at the Freie Universit\"at Berlin is gratefully acknowledged. Travel funds by the Max Planck Society via the International Max Planck Research School are appreciated.}


\begin{thebibliography}{10}

\bibitem{MIT1}
N.~F.~Mott, Rev.~Mod.~Phys. {\bf 40}, 677--683 (1968).

\bibitem{MIT2}
W.~Kohn, Phys.~Rev., {\bf 133}, A171 (1964).

\bibitem{MIT3}
E.~Lieb, and F.~Wu, Phys.~Rev.~Lett. {\bf 20}, 1445--1448 (1968).

\bibitem{MIT4}
M.~Imada, A.~Fujimori, and Y.~Tokura, Rev.~Mod.~Phys. {\bf 70}, 1039--1263 (1998).

\bibitem{MIT5}
R.~Resta, J.~Chem.~Phys., {\bf 124}, 104104 (2006).

\bibitem{Gebhard}
F.~Gebhard, \emph{The Mott Metal-Insulator Tansition: Models and Methods} (Springer, New York, 1997).

\bibitem{Mott}    
N.~Mott, \emph{Metal-Insulator Transitions}, 2nd ed. (Taylor \& Francis, London, 1990).

\bibitem{DFT1}    
R.~M.~Dreizler, E.~K.~U.~Gross, \emph{Density Functional Theory} (Springer, Berlin, 1990).

\bibitem{DFT2}    
H.~Eschrig, \emph{The Fundamentals of Density Functional Theory} (B. C. Teubner, Stuttgart, 1996).

\bibitem{DFT3}
W.~Kohn, L.~J.~Sham, Phys.~Rev.~A {\bf 140}, 1133 (1965).

\bibitem{paulus2003}
B.~Paulus, Chem. Phys. Lett. {\bf 371}, 7--14 (2003).

\bibitem{paulus2006}
B.~Paulus, Phys. Rep. {\bf 428}, 1--52 (2006).

\bibitem{paulus2007}
E.~Voloshina, and B.~Paulus, Phys. Rev. {\bf 75}, 245117 (2007).

\bibitem{stoll2009}
H.~Stoll, B.~Paulus, and P.~Fulde, Chem. Phys. Lett. {\bf 469}, 90--93 (2009).

\bibitem{Staemmler2009}
I.~Schmitt, K.~Fink, and V.~Staemmler, Phys. Chem. Chem. Phys. {\bf 11}, 11196--11206 (2009).

\bibitem{mueller2011}
C.~M\"uller, D.~Usvyat, and H.~Stoll, Phys. Rev. B {\bf 83}, 245136 (2011).

\bibitem{voloshina2012}
E.~Voloshina, Phys. Rev. {\bf 85}, 045444 (2012).

\bibitem{voloshina2014}
E.~Voloshina, and B.~Paulus, J. Chem Theory Comput. {\bf 10}, 1698--1706 (2014).

\bibitem{dmrg_mit}
E.~Fertitta, B.~Paulus, G.~Barcza, and \"{O}.~Legeza, Phys. Rev. B {\bf 90}, 245129 (2014).

\bibitem{chan-rev}
G.~K.-L.~Chan, and S.~Sharma, Annu. Rev.~Phys.~Chem. {\bf 62},  465--481 (2011).

\bibitem{reiher-rev}
K.~H.~Marti, and M.~Reiher, Z.~Phys.~Chem. {\bf 224},  583--599  (2010).

\bibitem{yanai-rev}
Y.~Kurashige, Mol.~Phys. {\bf 112}, 1485--1494 (2014).

\bibitem{wooters-rev}
S.~Wouters and D.~Van~Neck, European Phys. J. D {\bf 68}, 272 (2014).

\bibitem{szalay-rev}
Sz. Szalay, M. Pfeffer, V. Murg, G. Barcza, F. Verstraete, R. Schneider, O. Legeza, arXiv:1412.5829 (2014).

\bibitem{white92}
S.~R.~White, Phys.~Rev.~Lett. {\bf 69},  2863--2866  (1992).

\bibitem{white93}
S.~R.~White, Phys.~Rev.~B {\bf 48} 10345 (1992). 

\bibitem{legeza2003}
\"O.~Legeza, and J.~S\'olyom, Phys.~Rev.~B {\bf 68}, 195116 (2003).

\bibitem{legeza2004}
\"O.~Legeza, and J.~S\'olyom, Phys.~Rev.~B {\bf 70}, 205118 (2004).

\bibitem{legeza2006-qpt}
\"O.~Legeza, and J.~S\'olyom, Phys.~Rev.~Lett. {\bf 96}, 116401 (2006).

\bibitem{rissler06}
J.~Rissler, R.~M.~Noack, and S.~R.~White, Chem.~Phys. {\bf 323}, 519 (2006).

\bibitem{barcza2010a}
G.~Barcza, \"O.~Legeza, K.~H.~Marti, and M.~Reiher, Phys.~Rev.~A {\bf 83}, 012508 (2011).

\bibitem{boguslawski2012b}
K.~Boguslawski, P.~Tecmer, \"O.~Legeza, and M.~Reiher, J.~Phys.~Chem.~Lett. {\bf 3}, 3129--3135 (2012).

\bibitem{boguslawski2013a}
K.~Boguslawski, P.~Tecmer, G.~Barcza, \"O.~Legeza, and M.~Reiher, J.~Chem.~Theory~Comput. {\bf 9} 2959-2973 (2013).

\bibitem{CCSD1}
F.~Coester, Nucl.~Phys. {\bf 7}, 421 (1958).

\bibitem{CCSD2}
F.~Coester, and H.~Kümmel, Nucl.~Phys. {\bf 17}, 477 (1960).

\bibitem{CCSD3}
H.~Kümmel, Nucl.~Phys. {\bf 22}, 177 (1961).

\bibitem{CCSD4}
J.~Cizek, Adv.~Chem.~Phys. {\bf 14}, 35 (1969).

\bibitem{ref_CASSCF}
B.~O.~Roos, and P.~R.~Taylor, Chem.~Phys. {\bf 48},  157--173  (1980).

\bibitem{MOLPRO}
H.~J.~Werner, P.~J.~Knowles, G.~Knizia, F.~R.~Manby, M.~Sch\"{u}tz {\emph et al.}, {\sc MOLPRO}, version 2012.1, a package of ab initio programs, 2012, {\sc URL}: http://www.molpro.net.

\bibitem{Boys}
M.~Foster, and S.~F.~Boys, Rev.~Mod.~Phys. {\bf 32}, 296 (1960).

\bibitem{cc_pVDZ}
T.~H.~Dunning, J.~Chem.~Phys. {\bf 90}, 1007 (1989).

\bibitem{crystal}
R.~Dovesi, V.~R. Saunders, C.~Roetti, R.~Orlando, C.~M.~Zicovich-Wilson, F.~Pascale, B.~Civalleri, K.~Doll, N.~M.~Harrison, I.~J.~Bush, P.~D'Arco, and M.~Llunell,
CRYSTAL09 User's Manual (University of Torino, Torino, 2009). 

\end{thebibliography}
\end{document}